\documentclass[prb,superscriptaddress]{revtex4}
\usepackage{amsmath}
\usepackage{bm}
\usepackage{amssymb}
\usepackage{graphicx}
\usepackage[usenames,dvipsnames]{xcolor}

\def\bsigma{\bar\sigma}
\newcommand{\lr}{\langle}
\newcommand{\rr}{\rangle}

\newcommand{\cP}{{\cal P}}
\begin{document}
\title{Multiscale Cyclic Dynamics in Light Harvesting Complex in presence of vibrations and noise}
\author{Shmuel Gurvitz}
\email{shmuel.gurvitz@weizmann.ac.il}
\affiliation{Department of Particle Physics and Astrophysics, Weizmann Institute,
76100, Rehovot, Israel}
\affiliation{
Theoretical Division, T-4, Los Alamos National Laboratory, Los Alamos, NM, 87544, USA}
\author{Gennady P.  Berman}
\email{gpb@lanl.gov}
\affiliation{
Theoretical Division, T-4, Los Alamos National Laboratory, Los Alamos, NM, 87544, USA}
\author{Richard T. Sayre}
\email{rsayre@newmexiconsortium.org}
\affiliation{New Mexico Consortium, Los Alamos, NM, 87544, USA}

\begin{abstract}
Starting from the many-body Schr\"odinger equation, we derive a new type of Lindblad Master equations describing a cyclic exciton/electron dynamics in the light harvesting complex and the reaction center. These equations resemble the Master equations for the electric current in mesoscopic systems, and they go beyond the single-exciton description by accounting for the multi-exciton states accumulated in the antenna, as well as the charge-separation, fluorescence and photo-absorption. Although these effects take place on very different timescales, their inclusion is necessary for a consistent description of the exciton dynamics. Our approach reproduces both coherent and incoherent dynamics of exciton motion along the antenna in the presence of vibrational modes and noise. We applied our results to evaluate energy (exciton) and fluorescent currents as a function of sunlight intensity.
\end{abstract}

\maketitle

\section{Introduction}
\label{intro}
\begin{figure}[tbh]
\scalebox{0.3}{\includegraphics{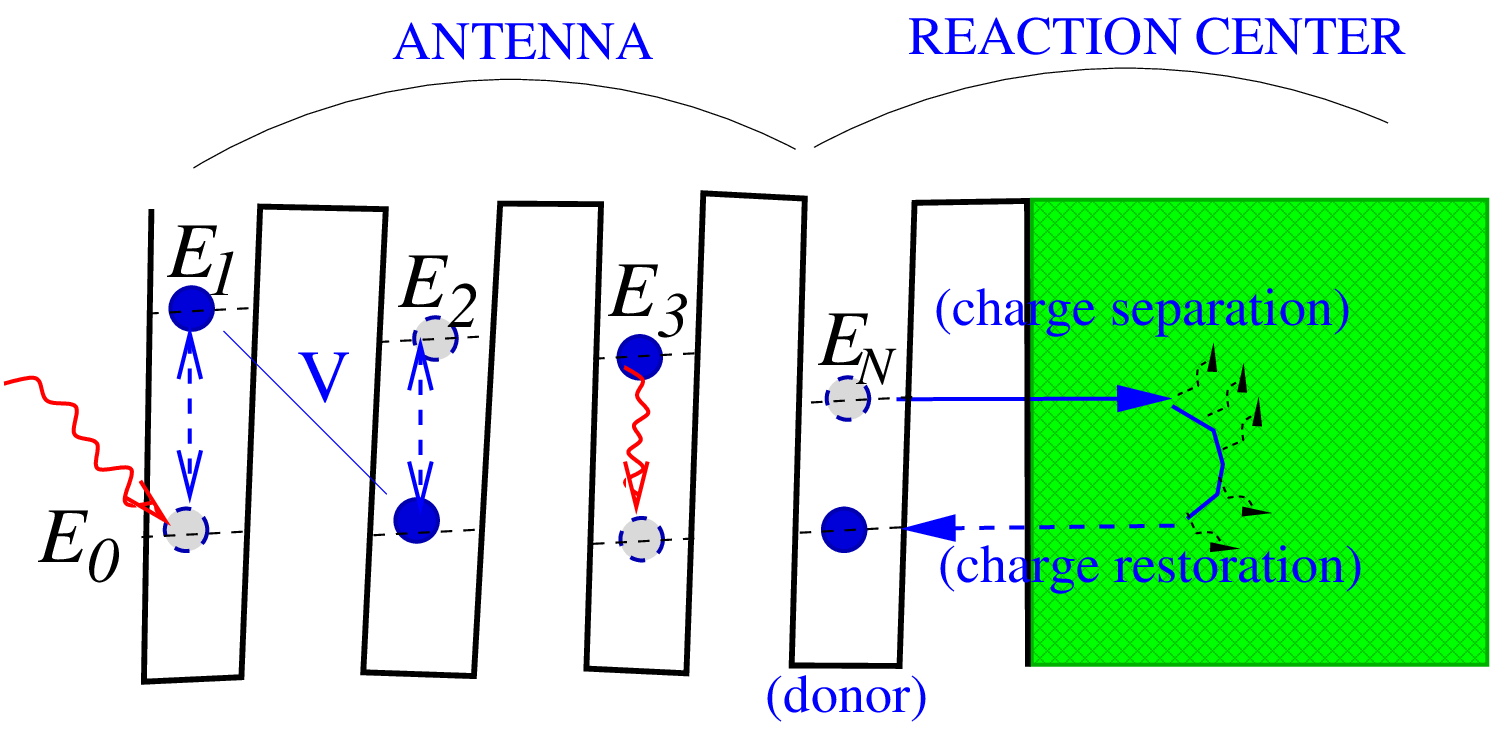}}$~~~~~~~~~~$ \scalebox{0.37}{\includegraphics{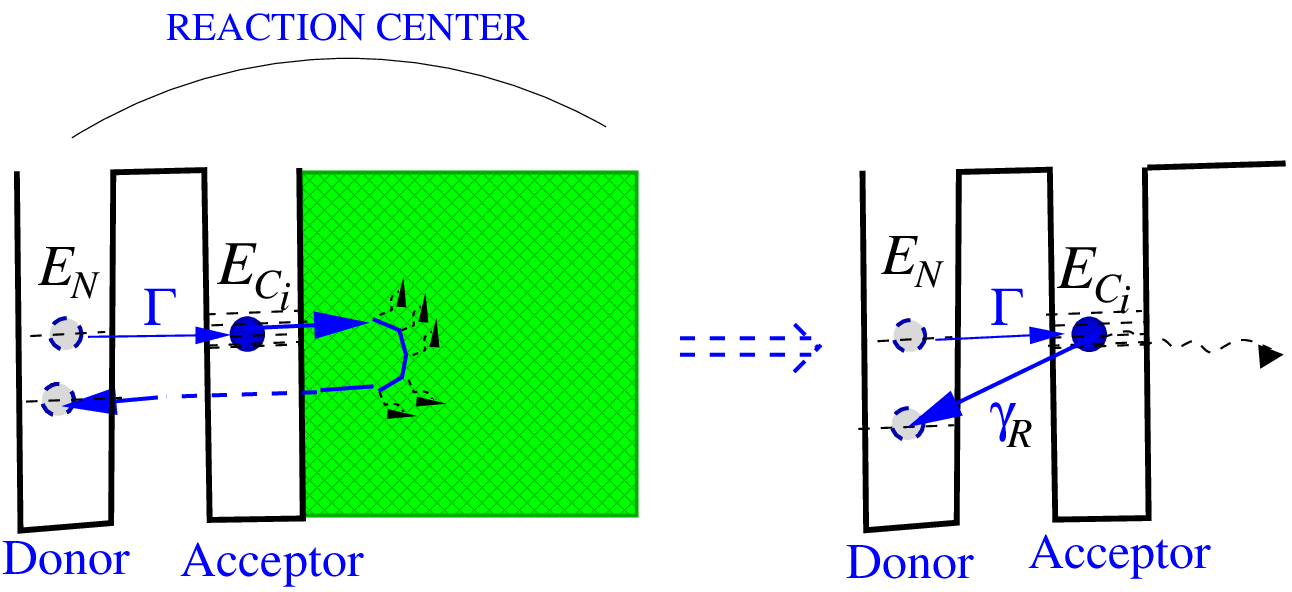}}
\caption{Left: schematic picture of the electron cycle. The cycle is completed with restoration of the donor's  neutrality (reduction). Right: the cycle is modeled by electron relaxation to the donor's ground state, through emission of the energy (fictitious boson) to the RC.}
\label{fig1}
\end{figure}

The energy transfer in the light-harvesting complex (LHC) takes place via exciton propagation among pigments bound to the LHC proteins \cite{book1}. The exciton is created by resonant photo-absorption in an antenna pigment, leading to electron excitation from the ground to the excited energy level, $\gamma +E_0 \to E_1$, Fig.~\ref{fig1}.  Due to the transitional dipole-dipole interaction, $V$, the exciton then propagates between neighboring pigments to the reaction center (RC), while all, excited and non-excited sites of the antenna, remain neutral. Finally, the exciton arrives at the site $N$ (the ``donor'' of the RC), where the primary charge separation occurs. The donor becomes positively charged,  and the electron participates in chemical reactions in the RC. Finally, the donor is neutralized (reduced) by an electron ultimately arrived from water splitting, Fig.~\ref{fig1}, and at some time, $\tau$, the cycle is completed.

The dynamics of the exciton transfer along the antenna, including the primary charge separation, is very rapid ($\sim$ ps). Otherwise, the exciton would be lost by fluorescence or by other (recombination) processes, taking place on the time-scale of $\sim$ ns. In comparison, the duration of entire cycle ($\tau$), completed with reduction of the oxidized primary electron donor, is much longer ($\sim \mu s$). During the cycle, no excitons occupy the RC donor. However, they can be accumulated by the antenna pigments, and finally being lost by fluorescence, as shown schematically in Fig.~\ref{fig1} (left panel).
Usually, at normal light intensity, the  study of energy-transfer in the LHC is limited by a single-exciton migration along the antenna pigment bed. At the same time, the research exists which goes beyond a single-exciton approach and which takes into account a cyclic regime and the multi-scale exciton dynamics in the LHCs (see for instance Refs.~\cite{Soler,Bruggemann}). However, the {\em consistent} quantum-mechanical consideration of the exciton-electron cyclic regime in antenna-reaction center, which includes: (i) photo-absorption, (ii) fluorescence in antenna, (iii) charge restoration of the RC donor and (iv) both coherent and incoherent exciton-electron dynamics, does not exist. This is mainly because all these effects occur at significantly different time-scales, and require the development of adequate quantum-mechanical mathematical approaches. Indeed, without a consistent accounting of all these multi-scale processes, in the frames of quantum consideration, one cannot fully understand and describe the exciton dynamics in the LHC \cite{book1}. In particular, it is related to accumulation of one or more excitons inside the antenna with increase of the light intensity. These ``trapped"  excitons can damage the photosynthetic apparatus through de-excitation pathways leading to generation of oxygen singlets and other damaging products.

At first sight, exciton transport along the LHC appears similar to spinless electron transport in a mesoscopic system. Indeed, no more than one exciton can reside on the same site, if only one excitation is allowed for each site (hard exciton model) \cite{muc}. As a result, the exciton propagation along the antenna would be similar to electron tunneling through coupled-dot system. The treatment of electron current through the coupled dots can be greatly simplified by reducing the many-body Schr\"odinger equation to the Lindblad-type particle-number-resolved Master equations \cite{gp,gur}. It is desirable to realize this analogy and derive similar Master equations for the exciton transport in the LHC, Fig.~\ref{fig1}. However, in this case we have to include restoration of the primary donor's neutrality, Fig.~\ref{fig1} (right panel), since a similar cycle dynamics is not considered in electron transport through coupled-dot systems.

Thus, the charge separation is effectively accounted for by coupling the donor's level $E_N$ to the acceptor, represented by a band of dense levels, $E_{C_i}$ (sink), as displayed in right panel of Fig.~\ref{fig1}. This results in irreversible tunneling of the electron from the donor to the sink with a rate, $\Gamma\sim$1/ps. In order to describe the effect of restoration of the donor's neutrality on the LHC dynamics, it is not necessary to know all details of the slow chemical reactions in the RC, initiated by the electrons. What is relevant, is a period of the cycle ($\tau$). Therefore, for our purposes, the RC can be considered as a ``Black Box'', absorbing the donor electron of the energy, $E_N$, and then emitting it (by time $\tau$) to the same donor site, but with a different energy, $E_0$.

This process can be modeled as a direct relaxation of an electron from the acceptor to the donor ground state, accompanied by emission of the energy ($E_N-E_0$) to the RC. Although the  cycle is completed by an electron coming from a different place (such as water splitting in the Photosystem II, etc.), its origin is not important  for the exciton dynamics in the LHC, in particular, since all electrons are indistinguishable. To account this relaxation
phenomenologically we add a fictitious field (boson bath) to the sink Hamiltonian, weakly coupled with all acceptor levels ($E_{C_i}$). If the bath is initially empty, the electron would exponentially decay to the donor's ground state with relaxation  decay rate, $\gamma_R=1/\tau$, by emitting fictitious bosons with energy, $E_{N}-E_0$, as displayed in Fig.~\ref{fig1} (right panel).

By modeling the energy transfer to the RC via quantized fictitious field, together with a quantized field, describing the light source and fluorescence, allows us to derive closed Master equations for exciton dynamics in a complete quantum mechanical way. As a result, we would be able to evaluate the energy (exciton) current and the fluorescent current as functions of the incoming sunlight intensity, and also the probability of single and multi-exciton states inside the antenna. This approach can be considered as a framework for the treatment of the energy transfer through any network in the antenna complex and also in the presence of vibrations and noise.

The paper is organized as follows. In Secs.~\ref{sec2} and \ref{sec21}, we describe the Master equation for the photo-absorption by a single excitonic site, as well as primary charge separation and restoration on the donor site by using our wave-function approach. Sec.~\ref{sec3} deals with the general case of an $N$-site antenna with a detailed example of the two-site antenna. Sec.~\ref{sec4} presents our account of the vibrational modes and the related dichotomic noise, generated by the environment. Last section is the Summary.

\section{Rate equations for photo-absorption\label{sec2}}

In order to understand a general structure and origin of the  particle-resolved Master equations (presented in Sec.~\ref{sec3}), describing the cyclic dynamics of the LHC, and conditions, at which these equations can be justified, we consider separately the exciton creation on the peripheral site and its decay to the RC from the inner site of the LHC.

Let us consider the photo-absorption on the first (peripheral) site of antenna, Fig.~\ref{fig1} (left panel), separated from the rest of antenna. The corresponding Hamiltonian, $H_1$, can be written as,
\begin{equation}
H_1=E_0\hat a_{0}^\dagger \hat a_{0}^{}+E_1\hat a_1^\dagger \hat a_1^{}+\sum_k\omega_k\hat C_k^\dagger \hat C_k^{}+\sum_k\big(g_k\hat B_1^\dagger\hat C_k+H.c.\big),
\label{apb1}
\end{equation}
where $\hat a_1^\dagger(\hat a_1^{})$ is an electron creation (annihilation) operator for the excited state, $E_1$, and $\hat a_{0}^\dagger (\hat a_{0}^{})$ is the same for the ground state, $E_0$, (in following we take $E_0=0$), while $\hat C_k^\dagger (\hat C_k^{})$ is a photon creation (annihilation) operator. $\hat B_1^{\dagger}=\hat a_1^{\dagger}\hat a_{0}^{}$ denotes an exciton creation operator. The last term in (\ref{apb1}) describes the electron-photon interaction in the rotating-wave approximation.

Using a similar technique, as for derivation of the particle-resolve rate equations for electron transport through muli-dot systems \cite{gur}, we arrive to the following   particle number-resolved Master equations of a form,
\begin{subequations}
\label{phot1}
\begin{eqnarray}
\dot{\sigma}^{(p)}_{00}(t)&=& - \Gamma_{in} \sigma^{(p)}_{00}(t)+\Gamma_{out} \sigma^{(p-1)}_{11}(t)\;,
\label{phot1a}\\
\dot{\sigma}^{(p)}_{11}(t)&=&  \Gamma_{in} \sigma^{(p)}_{00}(t)-\Gamma_{out} \sigma^{(p)}_{11}(t).
\label{phot1b}
\end{eqnarray}
\end{subequations}
where, $\sigma_{00}^{(p)}(t)$ and $\sigma_{11}^{(p)}(t)$, are probabilities  of finding the electron in the ground state ($E_0$) and in the excited state, ($E_1$), Fig.~\ref{fig1} with $p$ photons emitted by time $t$. Respectively, $\sigma_{\alpha\alpha}^{}(t)=\sum_p\sigma_{\alpha\alpha}^{(p)}$ where $\alpha=0,1$, are total probabilities ($\sigma_{00}(t)+\sigma_{11}(t)=1$).
Here $\Gamma_{in}$ and $\Gamma_{out}$ are rates of a photo-absorption, leading to electron transition from the ground to the excited state (exciton creation), and of a photo-emission in a reverse process (exciton annihilation). We found that $\Gamma_{in}=\bar n\gamma$ and $\Gamma_{out}=(\bar n+1)\gamma$, where $\bar n=n(E_1)$ is a number of photons with energy $E_1$ (which we call below the ``light intensity'') and
$\gamma=2\pi g^2(E_1)\rho (E_1)$, with $\rho$ being density of photon states.

It follows from our derivation that the validity of Eqs.~(\ref{phot1}) is based on Markovian approximation \cite{book3} (band-width is larger than $\gamma$). In this case, the contribution from the Green's function poles dominates in the equation of motion, leading to Eqs.~(\ref{phot1}). However, if the band-width is very narrow (less than $\gamma$), these equations have to be modified. This problem will be discussed in a separate publication.

Equations~(\ref{phot1}) are identical to those describing the electron transport from the source to the drain through a single quantum dot, with $\Gamma_{in}$ and $\Gamma_{out}$ corresponding to the incoming and outgoing electron rates \cite{gp,gur}. Summing up these equations over $p$ and taking into account that $\sigma_{11}(t)=1-\sigma_{00}(t)$, one easily finds the following rate equation for $\sigma_{00}(t)$ which can be rewritten as one equation,
\begin{align}
\dot{\sigma}_{00}(t)=-(2\bar n+1)\gamma\sigma_{00}(t)+(\bar n+1)\gamma,
\label{apb9a}
\end{align}

In the steady state limit, $\dot\sigma_{00}(t\to\infty)\to 0$, so the ground state occupation, $\bar\sigma_{00}=\sigma_{00}(t\to\infty)$ is $\bar\sigma_{00}=(\bar n+1)/(2\bar n+1)$. If the photon bath is in the thermal equilibrium state, then
$\bar n=1/(e^{E_1/T}-1)$. As a result, the occupation of the ground state is $\bar\sigma_{00}=1/(1+e^{-E_1/T})$,
which is a quite known result \cite{weiss}.

\section{Primary charge separation and restoration by emission of energy\label{sec21}}

Consider the site $N$ (``RC donor'') of antenna, coupled to  ``acceptor'', represented as a sink with dense levels, $E_{C_i}$, Fig.~\ref{fig1} (right panel). As a result, an electron, occupying the excited level ($E_N$) of the RC donor, tunnels to the ``acceptor'', leaving the donor positively charged (primary charge separation).  This process is very fast ($\sim$ ps) in a comparison with the time-scales of the subsequent chemical reactions in the RC ($\sim \mu s$). The cycle is completed, when the positively charged site, $N$, is neutralized (reduced) by an electron. This is modeled by a direct relaxation of the electron from the acceptor band, $E_{C_i}$, to the donor's ground state with emission of energy $E_N-E_0$ to the RC. The latter is represented by emission of a fictitious boson, carrying that energy. In order to describe these processes quantum-mechanically, we introduce an effective Hamiltonian, $H_N$, for the donor, $N$, Fig.~\ref{fig1} (right panel).
\begin{eqnarray}
H_N=E_N^{}\hat a_N^\dagger \hat a_N^{}+\sum\limits_i E_{C_i}^{}\hat a_{C_i}^\dagger\hat a_{C_i}^{}+\sum\limits_p\bar\omega_p^{}\hat F_p^\dagger \hat F_p^{}
+\sum\limits_i\Big(\tilde V_i^{}\hat a_{C_i}^\dagger \hat a_N^{}
+\sum\limits_p f_{ip}^{}\hat a_{0}^{\dagger}\hat a_{C_i}\hat F_p^{\dagger}+H.c.\Big).
\label{ham1}
\end{eqnarray}
Here, $\hat a_N^\dagger$ and $\hat a_{C_i}^\dagger$, denote electron creation operators at the site, $N$, and at a sub-level ($i$) of the acceptor,  $C$. Respectively, $\hat a_{0}^{\dagger}\equiv \hat a_{0N}^{\dagger}$, is an electron creation operator at the ground state of the donor (we chose $E_0\equiv E_{0N}=0$). The operator, $\hat F_p^{\dagger}$, describes a creation of  fictitious bosons, bearing the energy transferred to the RC and $\tilde V_i^{}$ is a tunneling coupling between the donor, $E_N$, with the sub-level, $E_{C_i}$, of the acceptor, and $f_{ip}^{}$ is a coupling of an electron on the acceptor with fictitious bosons.

Consider the one-electron cycle, completed with emission of one fictitious boson, displayed in right panel of Fig.~\ref{fig1}. The wave function, describing a whole system (electron and fictitious bosons) can be written as,
\begin{align}
|\Psi (t)\rangle &=\Big[b_N^{}(t)\hat a_N^\dagger a_{0}^{}+\sum_ib_{C_i}^{}(t)\hat a_{C_i}^\dagger\hat a_{0}^{}+\sum_p b_{0p}^{}(t)\hat F_p^\dagger\Big]|\bar 0\rangle,
\label{wf0}
\end{align}
where $|\bar 0\rangle\equiv \hat a_{0}^{\dagger}|0\rangle$ is the initial (``vacuum'') state of the system, corresponding to empty boson bath and the electron, occupying the donor's ground state.

Substituting Eq.~(\ref{wf0}) into the Schr\"odinger equation, $i\partial_t|\Psi (t)\rangle =H_N |\Psi (t)\rangle$, we find the system of coupled equations for the amplitudes $b(t)$ with the initial conditions: $b_N^{}(0)=1$ and $b_{C_i}^{}(0)=b_{0p}^{}(0)=0$. Using the same technique as in Ref.~\cite{gur}, we convert these equations to the following Master equations for the density matrix of the system, $\sigma_{NN}^{}(t)=|b_{N}^{}(t)|^{2},~~
\sigma_{CC}^{}(t)=\sum_i|b_{C_i}^{}(t)|^{2},
~~\sigma_{00}^{}(t)=\sum_p|b_{0p}^{}(t)|^2$,
\begin{subequations}
\label{eq3}
\begin{eqnarray}
\dot\sigma_{NN}^{}(t)&=&-\Gamma \sigma_{NN}^{}(t),\label{eq3a}\\
\dot\sigma_{CC}^{}(t) &=&\Gamma\sigma_{NN}^{}(t)-\gamma_R^{}\sigma_{CC}^{}(t),\label{eq3b}\\
\dot\sigma_{00}^{}(t)&=&\gamma_R^{}\sigma_{CC}^{}(t).\label{eq3c}
\end{eqnarray}
\end{subequations}
where, $\Gamma=2\pi|\tilde V|^2\rho_C^{}$ is the charge-separation rate and $\gamma_R^{}=(2\pi)^2|f|^2\rho_C\bar\rho\Delta=1/\tau$, is the rate of an entire cycle, with $\rho_C,\bar\rho$ are the density of states. Here, $\Delta\simeq\Gamma$, is a width of the electron distribution on the acceptor. Both rates are phenomenological parameters, which are determined experimentally ($1/\Gamma\sim ps$ and $1/\gamma_R^{}\sim \mu s$-ms).

\section{Exciton transport in $N$-site antenna\label{sec3}}

\subsection{Master equations in general case.\label{sec31}}

Now we extend our treatment on the $N$-site antenna chain, coupled with the electromagnetic field, describing photo-absorption and fluorescence, and with fictitious boson bath, describing the donor charge restoration,  Fig.~\ref{fig1}. The total Hamiltonian, describing this system is a combination of Eqs.~(\ref{apb1}), (\ref{ham1}), and it can be written as,
\begin{eqnarray}
{\cal H}_N&=\sum\limits_k\omega_k\hat C_k^\dagger \hat C_k^{}+\sum\limits_{m=1}^{N}E_m\hat B_m^\dagger \hat B_m^{}+\sum\limits_i E_{C_i}^{}\hat a_{C_i}^\dagger\hat a_{C_i}^{}
+\sum\limits_p\bar\omega_p\hat F_p^\dagger \hat F_p^{}+H_{int},
\label{ham}
\end{eqnarray}
where $\hat B_m^\dagger =\hat a_m^\dagger \hat a_{0m}^{}$ is an exciton creation operator on the site $m$. Here too we assume that the ground state energy for all sites $m=1,\ldots N$ is zero. All notations are the same as in Eqs.~(\ref{apb1}), (\ref{ham1}).
The interaction  term can be written as,
\begin{eqnarray}
&H_{int}=\sum\limits_{m=1}^N\sum\limits_k g_k^{}\,\hat B_m^\dagger \hat C_k^{}
+\sum\limits_{m=1}^{N-1} V_m^{}\,\hat B_{m+1}^\dagger \hat B_{m}^{}
+\sum\limits_i\Big(\tilde V_i^{}\hat a_{C_i}^\dagger \hat a_N^{}
+\sum\limits_p f_{ip}^{}\hat a_{0}^{\dagger}\hat a_{C_i}^{}\hat F_p^{\dagger}\Big)+H.c.
\nonumber
\end{eqnarray}
Here the electromagnetic field is coupled with all sites of antenna. However, excitons can be generated only on the first antenna site, $m=1$, by photon absorption. All other sites, $m=2,\ldots N$, are coupled with the empty photon reservoirs. Thus, the excitons occupying these sites can only decay by the fluorescence.

Note that the exciton commutation relations \cite{muc}, $
[\hat B_m^{},\hat B_n^\dagger]=\delta_{mn}^{}(1-2\hat B_m^\dagger \hat B_n^{})$, guarantee that two or more excitons cannot occupy the same site. Therefore, the exciton motion along the antenna, describing by the Hamiltonian~(\ref{ham}), is similar to that of the spinless electron transport trough the coupled quantum dots. The corresponding Master equation of the Lindblad-type can be derived from the time-dependent multi-particle Schr\"odinger equation, as discussed in previous examples. It represents a natural extension of Eqs.~(\ref{phot1}) and (\ref{eq3}), and can be written as (see Eq.~(75) of Ref.~\cite{gur}):
\begin{eqnarray}
\dot\sigma_{\alpha\alpha'}^{(\nu,\ell)} =  i\big({\cal E}_{\alpha'} - {\cal E}_{\alpha}\big) \sigma_{\alpha\alpha'}^{(\nu,\ell)}+i\sum_{\beta}\Big (\sigma_{\alpha\beta}^{(\nu,\ell)}
V_{\beta\to\alpha'}
-V_{\beta\to\alpha}\sigma_{\beta\alpha'}^{(\nu,\ell)}\Big)
-{1\over2}\sigma_{\alpha\alpha'}^{(\nu,\ell)}\sum_\beta
(\Gamma_{\alpha\to\beta}+\Gamma_{\alpha'\to\beta})
+\sum_{\beta,\beta'}\sigma_{\beta\beta'}^{(\nu',\ell')}
\Gamma_{\beta\to\alpha,\beta'\to\alpha'}
\label{d1}
\end{eqnarray}
where $|\alpha\rangle$, $|\beta\rangle$ enumerate all {\em discrete} multi-exciton states in the occupation number representation, and ${\cal E}_\alpha=\sum_{m\in\alpha}E_m$ is a total energy of the state, $|\alpha\rangle$.
The upper indices, $\nu$ and $\ell$, in the density matrix, $\sigma_{\alpha\alpha'}^{(\nu,\ell)}(t)$, denote the numbers of fluorescent photons and fictitious bosons emitted at time, $t$. Note, that in the last
(``gain'') term, $(\nu',\ell')=(\nu-1,\ell)$ or $(\nu',\ell')=(\nu,\ell-1)$, whenever emission of fluorescence photons or fictitious bosons takes place (c.f. with Eqs.~(\ref{phot1})).

The second term in Eq.~(\ref{d1}) describes the direct exciton transitions between neighboring sites, $V_{\beta\alpha}= V_{m,m+1}\equiv V_m$, via the transitional dipole-dipole interaction. One can realize that the first and second terms of Eq.~(\ref{d1}) represent the  commutator of the density matrix with the Hamiltonian in the Lindblad equation  \cite{lind}. The remaining two terms represent loss and gain processes generated by: (a) coupling of the site ($1$) to photon bath with rates  $\Gamma_{\alpha,\beta}\equiv\Gamma_{in,out}$, Eqs.~(\ref{phot1}) and all other sites with the rate $\Gamma_{\alpha,\beta}\equiv\gamma$; (b) charge separation with subsequent emission of fictitious bosons, leading to restoration of the donor's neutrality, with the rates $\Gamma_{\alpha,\beta}\equiv\Gamma,\gamma_R^{}$, respectively, Eqs.~(\ref{eq3}).

By solving Eqs.~(\ref{d1}), we can determine probabilities of any multi-exciton occupations, as well as the fluorescent current (in energy units), $I_{fl}(t)$, and the current of energy, $I_{en}(t)$, transferred to the RC. Those are given by (c.f. with Ref.~\cite{gp,gur}),
\begin{eqnarray}
I_{en}(t)&=&E_N\sum_{\nu,\ell}
\sum_{\alpha_N^{}}\ell\dot\sigma_{\alpha_N^{}\alpha_N^{}}^{(\nu,\ell)}(t)
=E_N^{}\gamma_R^{}\sum_{\alpha_N^{}}\sigma_{\alpha_N^{}\alpha_N^{}}^{}(t),
\nonumber\\
I_{fl}(t)&=&\sum_{\nu,\ell}
\sum_{m=2}^N\sum_{\alpha_m} E_{m}\nu\dot\sigma_{\alpha_m\alpha_m}^{(\nu,\ell)}(t)
=\gamma\sum_{m=2}^N\sum_{\alpha_m} E_{m}\sigma_{\alpha_m\alpha_m}^{}(t),
\label{cur3}
\end{eqnarray}
where the index,  $\alpha_m$ enumerates all multi-exciton states containing the site $m$, and $\sigma_{\alpha_m\alpha_m}(t)
=\sum_{\nu,\ell}\sigma_{\alpha_m\alpha_m}^{(\nu,\ell)}(t)$, is a corresponding occupation of these states, obtained from Eqs.~(\ref{d1}). Note, that the first antenna site, where an exciton is created, is excluded from the fluorescent current, Eq.~(\ref{cur3}). Respectively, the total fluorescent current is
$I_{flT}(t)=I_{fl}(t)+\Gamma_{out}E_1\sum_{\alpha_1^{}}
\sigma_{\alpha_1^{}\alpha_1^{}}(t)$, where
$\Gamma_{out}=(\bar n+1)\gamma$, Eq.~(\ref{phot1}).

\subsection{Two-site antenna.\label{sec32}}

\begin{figure}[h]\begin{center}
\scalebox{0.5}{\includegraphics{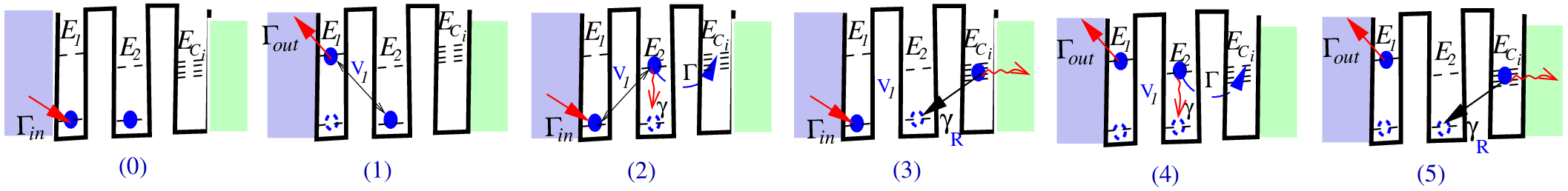}}
\caption{Exciton states of the two-site antenna. All allowed exciton transitions for each states are indicated.}
\label{fig5}\end{center}
\end{figure}

As an example for application of Eq.~(\ref{d1}), we consider exciton transport through the two-site antenna, ($N=2$). First, we need to enumerate all possible exciton states of the system, $\{\alpha,\beta\}=\{0,1\ldots, 5\}$. These are shown in Fig.~\ref{fig5}. Note, that the exciton propagates coherently between the sites (1) and (2) due to the transitional dipole-dipole interaction, $V_1$. All other transitions are incoherent, where the related transition rates are shown for each of states.

Now we can rewrite explicitly the Master equations~(\ref{d1}) for the reduced density-matrix, $\sigma_{\alpha\alpha'}\equiv \sigma_{\alpha\alpha'}(t)=\sum_{\nu,\ell}
\sigma_{\alpha\alpha'}^{(\nu,\ell)}(t)$, as
\begin{subequations}
\label{apb14}
\begin{eqnarray} \dot{\sigma}_{00}^{}&=&-\Gamma_{in}\sigma_{00}^{}
+\Gamma_{out}\sigma_{11}^{}
+\gamma\sigma_{22}^{}
+\gamma_R^{}\sigma_{33}^{}, \label{apb14a}\\
\dot{\sigma}_{11}^{}&=&i V_1(\sigma_{12}^{}-\sigma_{21}^{})
-\Gamma_{out}\sigma_{11}^{}
+\Gamma_{in}\sigma_{00}^{}
+\gamma\,\sigma_{44}^{}
+\gamma_R^{}\sigma_{55}^{},
\label{apb14b} \\
\dot{\sigma}_{22}^{}&=&i V_1(\sigma_{21}^{}-\sigma_{12}^{})
-(\gamma+\Gamma+\Gamma_{in})\sigma_{22}^{}
+\Gamma_{out}\sigma_{44}^{},
\label{apb14c} \\
\dot{\sigma}_{33}^{}&=&-(\Gamma_{in}
+\gamma_R^{})\sigma_{33}^{}
+\Gamma_{out}\sigma_{55}^{}
+\Gamma\sigma_{22}^{},
\label{apb14d}\\
\dot{\sigma}_{44}^{}&=&
-(\gamma+\Gamma_{out}+\Gamma)\sigma_{44}^{}
+\Gamma_{in}\sigma_{22}^{}, \label{apb14e}\\
\dot{\sigma}_{55}^{}&=&-(\Gamma_{out}
+\gamma_R^{})\sigma_{55}^{}
+\Gamma_{in}\sigma_{33}^{}
+\Gamma\sigma_{44}^{},\label{apb14f}\\
\dot{\sigma}_{12}^{}&=&i(E_{2}-E_1)\sigma_{12}^{}
+iV_1(\sigma_{11}^{}-\sigma_{22}^{})
-(\Gamma_T/2)\sigma_{12}^{}.
\label{apb14g}
\end{eqnarray}
\end{subequations}
where $\Gamma_T=\Gamma+\gamma+\Gamma_{in}
+\Gamma_{out}$. One can easily verify that these equations display the probability conservation, $\sum_{\alpha=0}^5\sigma_{\alpha\alpha}(t)=1$. Therefore it is useful to replace one of Eqs.~(\ref{apb14a})-(\ref{apb14f}) by the probability conservation.

Solving Eqs.~(\ref{apb14}), we find the energy (exciton) current to the RC, Eq.~(\ref{cur3})
\begin{align}
I_{en}(t)
=\gamma_R^{} E_2[\sigma_{33}(t)+\sigma_{55}(t)],
\label{cur33}
\end{align} whereas the fluorescent current (from the second site, Fig.~\ref{fig5}), and the total fluorescent current are given by

\begin{eqnarray}
I_{fl}(t)
&=&\gamma E_2[\sigma_{22}(t)+\sigma_{44}(t)]\, ,
\nonumber\\
I_{flT}(t)
&=&I_{fl}^{}(t)+\Gamma_{out}E_1[\sigma_{11}(t)
+\sigma_{44}(t)+\sigma_{55}(t)]\, .
\label{cur44}
\end{eqnarray}

\begin{figure}[h]
\scalebox{0.9}{\includegraphics{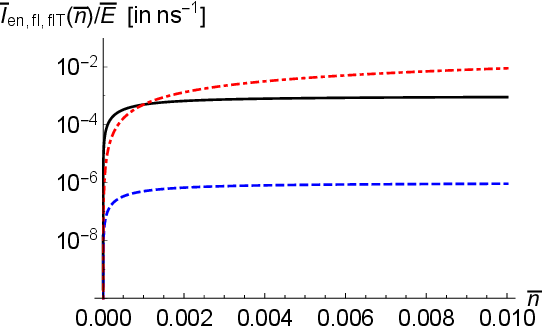}}$~~~~~~~~~$
\scalebox{0.9}{\includegraphics{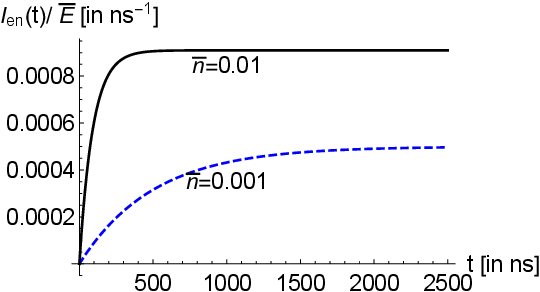}}
\caption{Left: the steady-state energy current, $\bar I_{en}$ (solid, black) and fluorescent current $\bar I_{fl}$ (dashed, blue) together with total fluorescent current (dot-dashed, red). Right: time-dependent energy current transferred to the RC for $\bar n=0.01$ (solid line, black) and $\bar n=0.001$ (dashed line, blue)}
\label{fig5pp}
\end{figure}

Consider now the steady-state limit, $t\to\infty$. Since in this limit  $\dot\sigma_{\alpha\alpha'}\to 0$, Eqs.~(\ref{apb14}) become a system of algebraic equations for $\bar\sigma\equiv \sigma (t\to\infty )$, which can be easily solved. The corresponding steady-state energy and fluorescent currents, Eqs.~(\ref{cur33}), (\ref{cur44}), are shown in Fig.~\ref{fig5pp} as functions of the light intensity ($\bar n$). Note, that $\bar n\gamma=\Gamma_{in}$, Eq.~(\ref{phot1}), is a number of photons absorbed by the first site per unit time. Here we chose for illustrative examples some generic values of parameters, not necessary related to a specific system, namely, $\gamma$=1/ns (fluorescent rate),  $\gamma_R=1/\tau=10^{-3}\gamma=1/\mu$s (charge restoration rate), $V_1=10^3\gamma$=1/ps (transitional dipole-dipole coupling between sites), and $\Gamma=10^3\gamma$=1/ps (charge separation rate).
The exciton energy levels of all sites are taken the same, $E_1=E_2=\bar E$.

The steady-state energy current to the RC, $\bar I_{en}$ (solid line, black) and the fluorescent current of the second site, $\bar I_{fl}$ (dashed line, blue) together with the total fluorescent current, $\bar I_{flT}$ (dot-dashed, red), divided by the donor energy $E_2=\bar E$, are shown in Fig.~\ref{fig5pp} in units of 1/ns, as functions  of $\bar n$. One finds from this figure that $\bar I_{en}$ and $\bar I_{fl}$ currents reach saturation already for a very small $\bar n$, where less than one photon is absorbed at the first site during one cycle ($\tau=1/\gamma_R=\mu$s). In addition, the fluorescent current $\bar I_{fl}$, Eq.~(\ref{cur44}), is found very small, in comparison with the energy current, $\bar I_{en}$, Eq.~(\ref{cur33}). This can be easily understood by taking into account that the energy current is the charge separation current. Indeed, it follows from Eqs.~(\ref{apb14}d,f) that $\gamma_R(\bar\sigma_{33}+\bar\sigma_{55})
=\Gamma(\bar\sigma_{22}+\bar\sigma_{44})$. Since the fluorescent current, $\bar I_{fl}$ comes from the donor site ($E_2$), where the charge separation  takes place, it proceeds with the rate $\Gamma$, which by 1000 larger than that the fluorescence rate from this site ($\gamma$). However, the total fluorescent current $\bar I_{flT}$, Eqs.~(\ref{cur44}), which include photon emission from the first site, is much larger than $\bar I_{fl}$ and exceeds the energy current with increase of the light intensity $\bar n$.

One also finds from the same figure (right panel) that time for approaching the asymptotic limit increases when the intensity of light ($\bar n$) decreases. It is interesting that for the light intensity $\bar n=0.001$ (one photon absorbed per a cycle), the current does not reach its steady-state value during the cycle, $t=\tau=1\mu$s.

In this paper, we restrict our consideration by calculating explicitly both the energy (exciton) current and the fluorescent current, for different values of parameters. A more detailed analysis of our results, including on the ET efficiency in the LHC, in the regime of the cyclic exciton dynamics, will be presented in a separate paper.

\section{Vibrational modes and noise\label{sec4}}

The role of vibrational modes in the exciton transport attracted recently much attention. It is now expected that  discrete sets of strongly coupled modes can considerably  enhance the transport properties along the antenna (see \cite{plenio} and references therein). The reason is that the near-resonant vibrations may effectively aligned the electron levels. We therefore consider only one vibrational state for each site, thus truncating the Hilbert space of vibrations by two states, $E_m$ and $E_m+\Omega$, where the vibrational frequency, $\Omega$, is assumed the same for all sites. All other vibrations are considered as a part of the environment, generating fluctuations between the two states of each site (c.f. with Ref. \cite{plenio}). This implies that the exciton transport along the antenna can be viewed as taking place through the time-dependent energy levels of the sites, $E_m\to E_m(t)=E_m+(\Omega /2)[1+ \xi_m(t)]$ in Eq.~(\ref{d1}),
where $\xi_m(t)=\pm 1$ is jumping randomly from 1 to -1 (or from -1 to 1) at a rate $\lambda_+$ (or $\lambda_-$), independently of its previous history. This represents so-called dichotomic or ``telegraph noise'', used in many models for fluctuating environment \cite{bergli,amnon,Gurv17}.

If the noise is generated by a heat bath of temperature $T$ (see for instance, Ref.~\cite{bergli}), then $\lambda_+^{}/\lambda_-^{}=
\bar \cP_-^{}/\bar \cP_+^{}=\exp[\Omega/(k_BT)]$, where $\bar \cP_\pm^{}$ are probabilities for finding $\xi_{m}(t)$ at the values $\xi_{m}=\pm 1$, ($\bar\cP_+^{}+\bar\cP_-^{}=1$).
The average value of $\xi_{m} (t)$ in the steady-state limit is therefore
\begin{align}
\bar\xi =\lr \xi_{m}(t)\rr=\sum_{\xi=\pm 1}\bar \cP_\xi \xi=(\lambda_-^{}-\lambda_+^{})/ \lambda=\big(1-e^{\Omega\over k_BT}\big)/\big(1+e^{\Omega\over k_BT}\big)
\label{rate2}
\end{align}
Thus, $\bar\xi=0$ for $T=\infty$, and $\bar\xi=-1$ for $T=0$.

Now we exemplify our procedure for the case of two-side antenna, considered in previous section. Let us average the density matrix $\sigma_{\alpha\alpha'}(t)$ directly in Eqs.~(\ref{apb14}), where $E_{1,2}$ in Eq.~(\ref{apb14g}) are replaced by $E_{1,2}+(\Omega /2)[1+ \xi_{1,2}(t)]$. One finds that Eqs.~(\ref{apb14a})-(\ref{apb14f}) keep the same form for the average density matrix $\lr\sigma_{\alpha\alpha'}(t)\rr$. The effect of noise appears only in Eq.~(\ref{apb14g}), which now reads
\begin{align}
&\lr\dot{\sigma}_{12}^{}\rr=i\epsilon\lr\sigma_{12}^{}\rr
+i(\Omega /2)(\lr\sigma_{12\xi_2}\rr
-\lr\sigma_{12\xi_1}\rr)
+iV_1(\lr\sigma_{11}^{}\rr-\lr\sigma_{22}^{}\rr)
-(\Gamma_T /2)\lr\sigma_{12}^{}\rr
\label{noise2}
\end{align}
where $\epsilon=E_2-E_1$, and
the terms $\lr\sigma_{12\xi_{1,2}}\rr\equiv \lr\xi_{1,2}(t)\sigma_{12}(t)\rr$  represent the effect of noise and vibrational modes in the equation of motion.  We consider the non-correlated noise acting on different sites, where $\lr\xi_1\xi_2\rr=\bar\xi^2$, Eq.~(\ref{rate2}).

In order to evaluate the second term in the r.h.s. of Eq.~(\ref{noise2}), we multiply each of Eqs.~(\ref{apb14}) by $\xi_{1,2}(t)$, taking into account that $\xi_{1,2}^2(t)=1$. For instance, multiplying Eq.~(\ref{apb14g}) by $\xi_2(t)$, we find
\begin{align}
&\lr\xi_2\dot{\sigma}_{12}^{}\rr=i\epsilon\lr\sigma_{12\xi_2}^{}\rr
+i{\Omega\over2}(1-\bar\xi^2)\lr\sigma_{12}\rr
+iV_1(\lr\sigma_{11\xi_2}^{}\rr-\lr\sigma_{22\xi_2}^{}\rr)
-{\Gamma_T \over2}\lr\sigma_{12\xi_2}^{}\rr\, .
\label{noise3}
\end{align}
This equation is still not useful, since its l.h.s. is not the  time-derivative of $\lr\sigma_{11\xi_2}^{}\rr$. However,
in the case of an exponential noise-correlator, $\lr\xi(t_1)\xi(t_1+\tau)\rr\propto\exp (-\lambda\, \tau)$, one can use the following very useful ``differential formula'', derived by Shapiro and Loginov \cite{shapiro,ga},
\begin{align}
{d\over dt}\langle \xi(t)R[\xi(t),t]\rangle&=\langle \xi(t){d\over dt}R[\xi(t),t]\rangle -\lambda\,\langle \xi(t) R[\xi(t),t]\rangle
+\lambda \bar\xi R[\xi(t),t]\, ,
\label{df}
\end{align}
where $\bar \xi$ is given by Eq.~(\ref{rate2}) and $R[\xi(t),t]$ is an arbitrary functional of the noise. In our case $R[\xi(t),t]\equiv \sigma_{12}(t)$, obtained from Eqs.~(\ref{apb14g}) with the time-dependent energy levels, $E_2-E_1\to\epsilon+(\Omega/2)[\xi_2(t)-\xi_1(t)]$. Substituting Eq.(\ref{df}) into Eq.~(\ref{noise3}) we find
\begin{align}
\lr\dot\sigma_{12\xi_2}^{}\rr=i\Big(\epsilon
+i{\Gamma_T+2\lambda\over2}\Big)\lr\sigma_{12\xi_2}^{}\rr
+\big[\lambda\bar\xi +i{\Omega\over2}(1-\bar\xi^2)\big]\lr\sigma_{12}\rr
+iV_1\big(\lr\sigma_{11\xi_2}^{}\rr
-\lr\sigma_{22\xi_2}^{}\rr\big)\, ,
\label{noise4}
\end{align}
Similar results are obtained for
Eqs.~(\ref{apb14a})-(\ref{apb14f}), by applying the differential formula~(\ref{df}). By proceeding in the same way with Eqs.~(\ref{apb14}), multiplied by the random variable $\xi_1(t)$, we finally obtain the closed set of the linear equations, describing the density matrix $\sigma_{\alpha\alpha'}(t)$ and the energy and fluorescent currents, Eqs.~(\ref{cur33}), (\ref{cur44}).

Consider the steady-state limit, $\bar\sigma_{\alpha\alpha'}=\lr\sigma_{\alpha\alpha'}(t\to\infty)\rr$, where the l.h.s. of Eqs.~(\ref{apb14}) vanishes. Then, these equations with Eq.~(\ref{noise2}) become algebraic for variables $\bar\sigma_{\alpha\alpha'}$ and $\bar\sigma_{12\xi_{1,2}}\equiv \langle\xi_{1,2}(t)\sigma_{12}(t)\rangle_{t\to\infty}$. The latter terms are obtained from Eqs.~(\ref{apb14}), multiplied by $\xi_{1,2}(t)$ in the limit $t\to\infty$. For instance, these equations for $\bar\sigma_{12\xi_{1}}$ become
\begin{subequations}
\label{nst14}
\begin{align}
&\bsigma_{00\xi_1}^{}+\bsigma_{11\xi_1}^{}+\bsigma_{22\xi_1}^{}
+\bsigma_{33\xi_1}^{}+\bsigma_{44\xi_1}^{}
+\bsigma_{55\xi_1}^{}=\bar\xi,\label{nst14a} \\
&i V_1(\bsigma_{12\xi_1}^{}-\bsigma_{21\xi_1}^{})
-\Gamma_{out}\bsigma_{11\xi_1}^{}
+\Gamma_{in}\bsigma_{00\xi_1}^{}
+\gamma\,\bsigma_{44\xi_1}^{}+\gamma_R^{}\bsigma_{55\xi_1}^{}
-\lambda\bsigma_{11\xi_1}+\lambda\bar\xi\bsigma_{11}
=0\, ,
\label{nst14b} \\
&i V_1(\bsigma_{21\xi_1}^{}-\bsigma_{12\xi_1}^{})
-(\gamma+\Gamma+\Gamma_{in})\bsigma_{22\xi_1}^{}
+\Gamma_{out}\bsigma_{44\xi_1}^{}
-\lambda\bsigma_{22\xi_1}+\lambda\bar\xi\bsigma_{22}
=0,
\label{nst14c} \\
&-(\Gamma_{in}+\gamma_R^{})\bsigma_{33\xi_1}^{}
+\Gamma_{out}\bsigma_{55\xi_1}^{}+\Gamma\bsigma_{22\xi_1}^{}
-\lambda\bsigma_{33\xi_1}+\lambda\bar\xi\bsigma_{33}
=0,
\label{nst14d}\\
&\-(\gamma+\Gamma_{out}+\Gamma)\bsigma_{44\xi_1}^{}
+\Gamma_{in}\bsigma_{22\xi_1}^{}
-\lambda\bsigma_{44\xi_1}+\lambda\bar\xi\bsigma_{44}
=0, \label{st14e}\\
&-(\Gamma_{out}+\gamma_R^{})\bsigma_{55\xi_1}^{}
+\Gamma_{in}\bsigma_{33\xi_1}^{}+\Gamma\bsigma_{44\xi_1}^{}
-\lambda\bsigma_{55\xi_1}+\lambda\bar\xi\bsigma_{55}
=0,
\label{nst14f}\\
&i\Big(\epsilon+i{\Gamma_T+2\lambda\over2}\Big)\bsigma_{12\xi_1}^{}
+\big[\lambda\bar\xi-i{\Omega\over2}(1-\bar\xi^2)\big]\bsigma_{12}
+iV_1(\bsigma_{11\xi_1}^{}-\bsigma_{22\xi_1}^{})=0.
\label{nst14g}
\end{align}
\end{subequations}
where Eq.~(\ref{nst14a}) is obtained from the probability conservation. Respectively, equations for $\bar\sigma_{12\xi_{2}}$ are the same with a replacement $\xi_1\to\xi_2$ and $\Omega\to -\Omega$ in Eq.~(\ref{nst14g}). By solving these equations, we can find the steady-state currents in the system, Eqs.~(\ref{cur33}), (\ref{cur44}).
\begin{figure}[h]
\scalebox{0.9}{\includegraphics{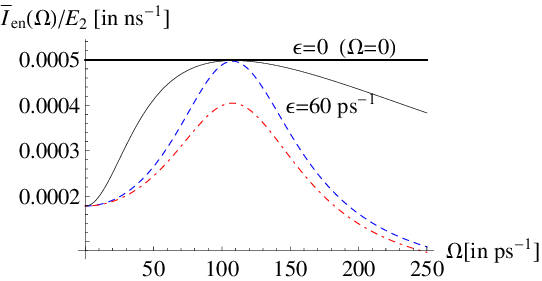}}$~~~~~~~$
\scalebox{0.8}{\includegraphics{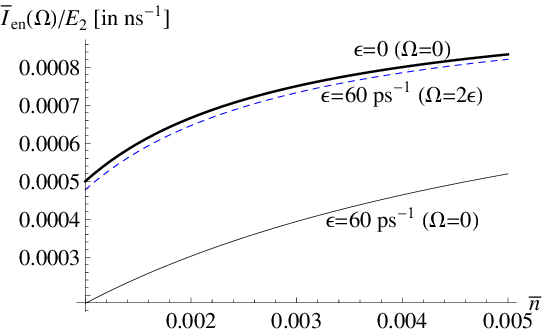}}
\caption{Energy current in 2-site antenna in the presence and without vibrations and noise.}
\label{fig5ppp}
\end{figure}

An example of such calculations is presented in Fig.~\ref{fig5ppp}. Left panel presents steady-state energy current, $I_{en}/E_2$ as a function of the vibration frequency, $\Omega$, for misaligned levels ($\epsilon=60$ps$^{-1}$) and the light intensity $\bar n=.001$, corresponding to one photon, absorbed by the first site per a cycle (1 $\mu$s). Three curves correspond to different values of the noise-spectrum width: $\lambda =0.001,\, 0.1,\, 10$ ps$^{-1}$ (dot-dashed red, dashed blue and solid (thin) lines, respectively). The noise is taken at room temperature.  For a comparison, thick solid line (black) shows the result for aligned levels, but without vibrations and noise. We should point out that the effect of noise disappears in the limit $\lambda\to 0$ and also for $\lambda\to\infty$ (not shown here).

Right panel shows the energy current as a function of the light intensity for misaligned levels without the noise and vibrations (thin solid line black) and with the noise at the resonance ($\Omega=2\epsilon$, dashed line blue). Thick solid line (black) displays the result for aligned levels without the noise. The results shown in Fig.~\ref{fig5ppp} demonstrate that incoherent fluctuations between vibronic levels can greatly increase the energy current without stringent resonance condition.

\section{Summary\label{sec5}}

Although most investigations of energy (exciton) transport in the LHCs concentrate on one-exciton motion along the antenna, without inclusion of a very slow cyclic dynamics,  we demonstrated that it is not sufficient for a consistent description of exciton dynamics. Therefore, we extended the Hamiltonian by including additional parts, describing the exciton creation and the fluorescence, through the interaction with the electromagnetic field, charge separation on the donor site, and the charge restoration after completing the corresponding cycle of the chemical reactions in the RC. The latter part is described phenomenologically, as an electron relaxation from the RC to the donor's ground state by emission of fictitious bosons, representing the energy transfer to the RC.

Since our cycled Master equations represent a more detailed description of the LHC dynamics, one can evaluate important effects, which cannot be treated by other methods. For instance, despite a very rapid exciton transfer along the antenna, $\sim$ ps, a very slow cyclic dynamics strongly affects the exciton current. This is unavoidable, since the current does not reach its steady value during a cycle. Moreover, the slow cycle dynamics enhances very drastically the loss of exciton due to fluorescence (taking place on the scale of $\sim$ ns). The reason is that the excitons are trapped during the cycle, and even one exciton trapping decreases the efficiency very strongly. The multi-exciton trapping, appearing at stronger light intensity, increases the fluorescent current furthermore. These states can also be relevant for artificial photosynthetic systems.

The influence of vibrational structures on exciton transport in LHC  attracted considerable attention last years.  Indeed, near-resonance under-damped vibrations can considerably enhance transport properties. This assumes, however, a fine adjustment  between coherent vibration frequency and a mismatch between the energy levels of neighboring  sites, carrying the exciton current. In this work, we considered a different approach, where  interaction with the thermal environment turns the coherent vibrations into incoherent fluctuations between the vibronic levels. As a result, the exciton transport along the antenna can be described by the (time-dependent) tunneling Hamiltonian, where the energy levels of each site are under the dichotomic (telegraph) noise, which is in the thermal equilibrium with the environment. We demonstrated here that the dichotomic noise can be treated exactly, without additional complications of the equations of motion.

\section*{Acknowledgment}
This work was carried out under the auspices of the National Nuclear Security
Administration of the U.S. Department of Energy at Los Alamos National Laboratory
under Contract No. DE-AC52-06NA25396.  R.T.S. acknowledges support from the LDRD program at LANL. S.G. is thankful to the CNLS of LANL for its hospitality and for a financial support of his visit, where a part of this work was done.

\section*{Author contributions statement} 
S.G. contributed in developing a mathematical approach and an illustrative material. G.P.B. contributed in modeling the LHC-RC system and in estimation of required parameters. D.T.S. contributed in adjusting parameters to the experimentally available.

\end{document}